\documentclass[12pt]{article}
\usepackage{graphics,cite,amssymb,epsfig,float,psfrag}
\usepackage[usenames,dvips]{color}
\usepackage{rotating}

\begin{document}

\newcommand{\lsim}{\raisebox{-0.13cm}{~\shortstack{$<$ \\[-0.07cm] $\sim$}}~}
\newcommand{\gsim}{\raisebox{-0.13cm}{~\shortstack{$>$ \\[-0.07cm] $\sim$}}~}
\newcommand{\dx}{\mbox{\rm d}}
\newcommand{\ra}{\rightarrow}
\newcommand{\lra}{\longrightarrow}
\newcommand{\ee}{e^+e^-}
\newcommand{\gam}{\gamma \gamma}
\newcommand{\tb}{\tan \beta}
\newcommand{\mlsp}{m_{\tilde \chi_1^0}}
\newcommand{\s}{\smallskip}
\newcommand{\nn}{\noindent}
\newcommand{\non}{\nonumber}
\newcommand{\beq}{\begin{eqnarray}}
\newcommand{\eeq}{\end{eqnarray}}
\newcommand{\pslash}{\not\hspace*{-1.6mm}p}
\newcommand{\kslash}{\not\hspace*{-1.6mm}k}
\newcommand{\lslash}{\not\hspace*{-1.6mm}l}
\newcommand{\eslash}{\hspace*{-1.4mm}\not\hspace*{-1.6mm}E}
\baselineskip=17pt
\thispagestyle{empty}

\hfill LPT--Orsay--05--25

\hfill April 2005

\vspace*{.5cm}

\begin{center}

{\sc\Large\bf Neutralino Dark Matter in mSUGRA:}

\vspace{0.5cm}

{\Large\sc\bf Reopening the light Higgs pole window}

\vspace{0.7cm}

{\sc\large Abdelhak Djouadi$^1$, Manuel Drees$^2$} and {\sc\large Jean-Loic 
Kneur$^3$}
 
\vspace*{7mm} 

$^1$ Laboratoire de Physique Th\'eorique d'Orsay, UMR8627--CNRS,\\
Universit\'e Paris--Sud, B\^at. 210, F--91405 Orsay Cedex, France. 
\vspace*{2mm}

$^2$ Physikalisches Institut, Universit\"at Bonn, \\
Nussallee 12, D--53115 Bonn, Germany. 
\vspace*{2mm}

$^3$ Laboratoire de Physique Th\'eorique et Astroparticules, UMR5207--CNRS,\\
Universit\'e de Montpellier II, F--34095 Montpellier Cedex 5, France. 

\end{center} 

\vspace*{5mm} 

\begin{abstract} 
  
\nn 

The requirement that the lightest neutralino $\tilde\chi_1^0$ has the right
thermal relic density to explain all Dark Matter in the universe strongly
constrains the parameter space of supersymmetric models in general, and of the
mSUGRA model in particular. Recently improved calculations of
the mass of the light CP--even Higgs boson $h$ present in this model, and the
increased central value of the mass of the top quark, have re--opened the
possibility that $2 \mlsp \lsim m_h$. In this ``$h-$pole region'' the LSP
annihilation cross section is enhanced by near--resonant $h$ exchange in the
$s-$channel, reducing the relic density to acceptable values. We delineate the
corresponding region of mSUGRA parameter space, and explore its phenomenology.
In particular, we find strong upper bounds on the masses of the gluino,
lighter chargino and LSP.

\end{abstract}

\newpage

\section{Introduction}

Supersymmetrizing the phenomenologically very successful Standard Model (SM)
of particle physics has many advantages. In addition to solving the (technical
aspect of) the hierarchy problem \cite{hier}, it is also compatible with the
Grand Unification of all gauge interactions \cite{gut}. In addition, if $R$
parity is conserved, the lightest superparticle (LSP) is stable, making it a
possible candidate for the cold dark matter (DM) in the universe, the
existence of which is inferred from cosmological observations \cite{dmrev}, in
particular from detailed analyses of the anisotropy of the cosmic microwave
background \cite{wmap}.

The necessary breaking of supersymmetry in general introduces many unknown
pa\-ra\-me\-ters. Most of these parameters are associated with flavor mixing
and/or CP--violation, and are severely constrained by experiment. This
motivates the study of models where additional sources of flavor changing
neutral currents (FCNC) and CP--violation are automatically suppressed. These
models have the additional advantage of being able to describe the entire
superparticle and Higgs spectrum with a small number of free parameters, which
dramatically increases their predictive power. The oldest such model
\cite{mSUGRA,nilles} goes under the name of minimal Supergravity (mSUGRA).
Here one assumes that gaugino masses, soft breaking scalar masses, and
trilinear soft breaking parameters all have universal values, $m_{1/2}, \,
m_0$ and $A_0$, respectively, at the scale of Grand Unification $M_X \simeq 2
\cdot 10^{16}$ GeV. Unlike models with gauge \cite{gmsb} or anomaly
\cite{amsb} mediated supersymmetry breaking, mSUGRA allows the lightest
neutralino $\tilde \chi_1^0$ as LSP in the visible sector to have the required
thermal relic density for natural masses (in the range of tens to hundreds of
GeV).

At least in the framework of standard cosmology, the Dark Matter density is by
now quite well known, in particular from the WMAP data \cite{wmap}. In our
analysis we will use the 99\%  (confidence level) CL region
\beq \label{omrange}
0.087 \leq \Omega_{\tilde \chi_1^0} h^2 \leq 0.138\, ,
\eeq
where $\Omega_{\tilde \chi_1^0}$ is the LSP mass density in units of the
critical density, and $h$ is today's Hubble constant in units of 100
km/(s$\cdot$Mpc).  Not surprisingly, this requirement greatly constrains the
allowed parameter space. Here we work under the usual assumption that the LSP
once was in thermal equilibrium; its relic density is then essentially
inversely proportional to its annihilation cross section \cite{dmrev}. Recent
analyses \cite{others,ddk1} found four distinct ``cosmologically acceptable''
regions\footnote{An additional region, with co-annihilation of the LSP with 
top squarks \cite{stop-co} is in general disfavored in mSUGRA--type scenarios.}
 where these assumptions lead to a relic density in the range
(\ref{omrange}).  Scenarios where both $m_0$ and $m_{1/2}$ are rather small
(the ``bulk region'') are most natural from the point of view of electroweak
symmetry breaking, but are severely squeezed by lower bounds from searches for
superparticles and Higgs bosons \cite{pdg}. In the ``co--annihilation'' region
one has $\mlsp \simeq m_{\tilde \tau_1}$, leading to enhanced destruction of
superparticles since the $\tilde \tau_1$ annihilation cross section is about
ten times larger than that of the LSP; this requires $m_{1/2} \gg m_0$. The
``focus point'' or ``hyperbolical branch'' region occurs at $m_0 \gg m_{1/2}$,
and allows $\tilde \chi_1^0$ to have a significant higgsino component,
enhancing its annihilation cross sections into final states containing gauge
and/or Higgs bosons; however, if $m_t$ is near its current central value
\cite{d0top} of 178 GeV, this solution requires multi--TeV scalar masses.
Finally, if the ratio of vacuum expectation values $\tan\beta$ is large, the
$s-$channel exchange of the CP--odd Higgs boson $A$ can become nearly
resonant, again leading to an acceptable relic density (the ``$A-$pole''
region).

Here we emphasize that a fifth cosmologically acceptable region of mSUGRA
parameter space exists. In a significant region of parameter space one has $2
\mlsp \lsim m_h$, so that $s-$channel $h$ exchange is nearly resonant. This
``$h-$pole'' region featured prominently in early discussions of the Dark
Matter density in mSUGRA \cite{old}, but seemed to be all but excluded by the
combination of rising lower bounds on $m_h$ and\footnote{In the relevant
  region of parameter space, $\mlsp \simeq m_{\tilde \chi_1^\pm}/2$; lower
  bounds on the chargino mass therefore directly translate into lower bounds
  on the LSP mass.} $\mlsp$ from searches at LEP \cite{pdg}. However, in
recent years improved calculations \cite{newhiggs} of the mass of the light
CP--even $h$ boson and the increase of the central value of the top mass
to 178 GeV \cite{d0top} have resurrected\footnote{Ref.\cite{baer1}, which uses
  $m_t = 175$ GeV, finds a small $h-$pole region for $\tan\beta=10$, but not
  for larger values of $\tan\beta$. Ref.\cite{baer1a}, which uses $m_t = 180$
  GeV, finds an $h-$pole region for $\tan\beta = 30$. Neither of these papers
  systematically analyzes the extent of this region.} this possibility;
improved calculations \cite{bagger} of the neutralino and chargino mass
spectrum also play a role. The relevant region of parameter space is
delineated in Sec.~2 and the resulting phenomenology is discussed in Sec.~3.
Sec.~4 contains a brief summary and some conclusions.


\vspace*{-2mm}
\section{The $h-$pole region}

The mSUGRA parameter space is defined by four continuous parameters and one
sign,
\beq \label{params}
m_0, \, m_{1/2}, \, A_0, \, \tan\beta, \, {\rm sign}(\mu).
\eeq
The common scalar soft breaking mass $m_0$, the common gaugino mass $m_{1/2}$
and the common trilinear soft term $A_0$ are all defined at the scale $M_X$
where the running ($\overline{\rm DR}$) $U(1)_Y$ and $SU(2)$ gauge couplings
meet. The ratio of vacuum expectation values $\tan\beta$ is defined at the
weak scale, which we take to be the geometric mean of the masses of the two
$\tilde t$ mass eigenstates. The sign of the supersymmetric higgs(ino) mass
parameter $\mu$ is independent of the scale. The connection between the weak
scale and $M_X$ is established by a set of coupled renormalization group
equations (RGE) \cite{rge}. The RG evolution can drive one combination of
squared Higgs boson masses to negative values, thereby triggering electroweak
symmetry breaking (EWSB) \cite{radbreak}. The requirement that this leads to
the correct mass of the $Z$ boson fixes the absolute value of $\mu$. 

We use the Fortran package SUSPECT \cite{suspect} to solve the RGE and to
calculate the spectrum of physical sparticles and Higgs bosons, following the
procedure outlined in \cite{ddk1}. Of special interest to the present study is
that this calculation includes leading one--loop ``threshold'' corrections to
neutralino and chargino masses as well as two--loop corrections to the
corresponding RGE. Also included are complete one--loop corrections and many
two--loop corrections \cite{ben} to scalar Higgs boson masses. As mentioned
in the Introduction, some of these corrections have only been calculated in
the last few years \cite{newhiggs}; they increase the mass $m_h$ of the light
CP--even Higgs boson $h$ by several GeV.

In addition to leading to consistent EWSB, a given set of input parameters has
to satisfy several experimental constraints. The ones relevant for this study
are: \begin{itemize}
\vspace*{-2mm}

\item The total cross section for the production of any pair of superparticles
  at the highest LEP energy (206 GeV) must be less than 20 fb. The only
  exception is the $\tilde \chi_1^0 \tilde \chi_1^0$ final state, which is
  invisible.\footnote{Note that we assume $R-$parity to be conserved, and that
  the gravitino mass is larger than that of the lightest sparticle in the
  visible sector.} This is a rather aggressive interpretation of LEP
  sparticle search limits \cite{lepsusy,pdg}; the true limits are often
  somewhat weaker than this. However, since the cross sections near threshold
  increase very quickly with decreasing sparticle masses, the resulting limits
  match experimental results rather closely in most cases.
\vspace*{-2mm}
  
\item Searches for neutral Higgs bosons at LEP \cite{lephiggs,pdg} impose a
  lower bound on $m_h$; note that in the region of mSUGRA parameter space of
  interest here, the couplings of $h$ to SM particles are very similar to that
  of the single Higgs boson of the SM. Allowing for a theoretical uncertainty
  \cite{herror} in the calculation of $m_h$ of about 3 GeV, we therefore
  require the calculated value of $m_h$ to exceed 111 GeV.
\vspace*{-2mm}
  
\item Recent measurements \cite{amuexp} of the anomalous magnetic moment of
  the muon lead to the constraint on the supersymmetric contribution
  \cite{amususy} to $a_\mu \equiv (g_\mu-2)/2$
\beq \label{amu}
-5.7 \cdot 10^{-10} \leq a_{\mu,\, {\rm SUSY}} \leq 4.7 \cdot 10^{-9}.
\eeq
  This range has been constructed from the overlap of the 2$\sigma$ allowed
  regions using data from $e^+e^-$ annihilation into hadrons and from $\tau$
  decays, respectively, to estimate the (hadronic) SM contribution to $a_\mu$;
  see refs.\cite{amuth} for discussions of this theoretical uncertainty.
\vspace*{-2mm}
  
\item The calculated $\tilde \chi_1^0$ relic density has to be in the range
  (\ref{omrange}). Our calculation uses proper thermal averaging of the
  squared $s-$channel, in particular $h-$exchange, contribution near the
  resonance \cite{gs}, while all other contributions are treated using the
  standard non--relativistic expansion. Note that the total $\tilde \chi_1^0$
  annihilation cross section is completely dominated by $h-$exchange diagrams
  in the region of parameter space of interest to this analysis.
\vspace*{-2mm}

\item Allowing for experimental and theoretical errors \cite{pdg}, the
  branching ratio for radiative $b$ decays should satisfy
\beq \label{bsg}
2.65 \cdot 10^{-4} \leq B(b \rightarrow s \gamma) \leq 4.45 \cdot 10^{-4}.
\eeq
  We evaluate this branching ratio, including contributions from $t H^\pm$ and
  $\tilde t \tilde \chi^\pm$ loops, using the results of ref.\cite{gambino}.
\end{itemize}

We consider this last constraint to be not as reliable as the other limits
discussed above. First of all, we are not aware of any complete fit of the
elements of the quark mixing (Kobayashi--Maskawa) matrix in the context of
mSUGRA (or any other extension of the SM). We follow the usual practice of
using the value of $V_{ts}$ as extracted in the framework of the SM when
evaluating the branching ratio; note that $B(b \rightarrow s \gamma) \propto
|V_{ts}|^2$. Moreover, this calculation would be affected significantly if one
allowed small deviations from universality, or equivalently, small
non--diagonal entries in the squark mass matrix, at the input scale
\cite{sqmix}. This modification would leave all other results, including the
sparticle spectrum and the LSP relic density, nearly unchanged. 

\begin{figure}[!h]
{ \hspace*{1.5cm} $\mathbf{m_0}$ [GeV]} 
\begin{center}
\mbox{\epsfig{file=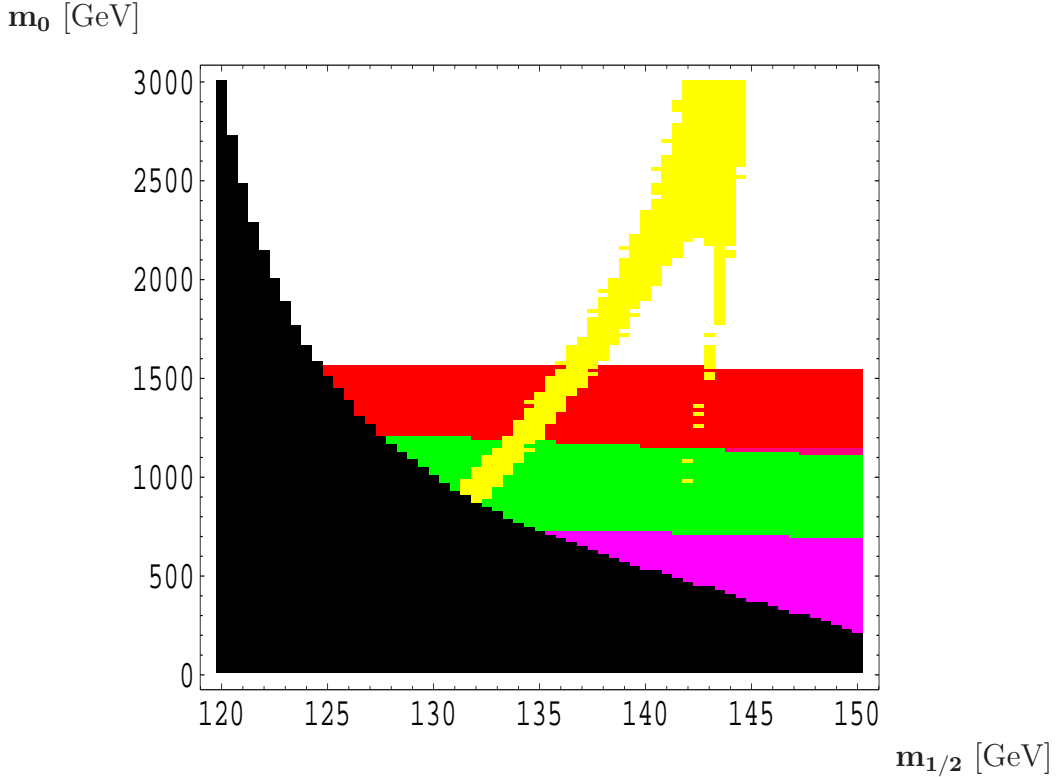,width=10cm,height=9cm}}
\end{center}
\vspace*{-.5cm}
{ \hspace*{13.3cm}  $\mathbf{m_{1/2}}$ [GeV]}

\caption[]{Constraints on the $m_{1/2}$--$m_0$ mSUGRA parameter space for 
  $m_t=178$ GeV, $A_0=0$, $\tan\beta=30$ and $\mu>0$. The dark area is ruled
  out by the requirement of EWSB and sparticle search limits, as discussed in
  the text.  The violet and green areas are ruled out by, respectively, the
  LEP Higgs search and the $b \to s\gamma$ constraints. The red area
  corresponds to the LEP evidence for a light MSSM Higgs boson with $m_h \sim
  116$ GeV. In the yellow bands the neutralino relic density falls in the
  range favored by WMAP, $0.087 \leq \Omega_{\tilde \chi_1^0} h^2 \leq
  0.138$.}  \vspace*{-.1cm}
\end{figure}

We are now ready to present some numerical results. In Fig.~1 we show the
relevant region of the $(m_0, m_{1/2})$ plane for $A_0 = 0, \, \tan\beta = 30$
and $m_t = 178$ GeV. The region $m_0 < 700$ GeV is excluded by Higgs searches
at LEP. If the theoretical uncertainty of the calculation of $m_h$ is ignored,
i.e. if we require the calculated $m_h$ to exceed 114 GeV, values $m_0 < 1150$
GeV would be excluded; this is coincidentally very close to the limit on $m_0$
from the constraint on $b \rightarrow s \gamma$ decays (green area). The
(weak) evidence for an SM--like Higgs boson with mass $\sim 116$ GeV
\cite{lephiggs} favors the red region. Finally, in the yellow region the LSP
relic density satisfies (\ref{omrange}). This region extends to very large
$m_0$, and eventually merges with the focus point/hyperbolical branch region.
At smaller $m_0$ it splits into two bands, with $\Omega_{\tilde \chi_1^0}$
being too low between the two branches.

\begin{figure}[!ht]
\begin{center}
\vspace*{-2.3cm}
\hspace*{-2.2cm}
\mbox{\epsfig{file=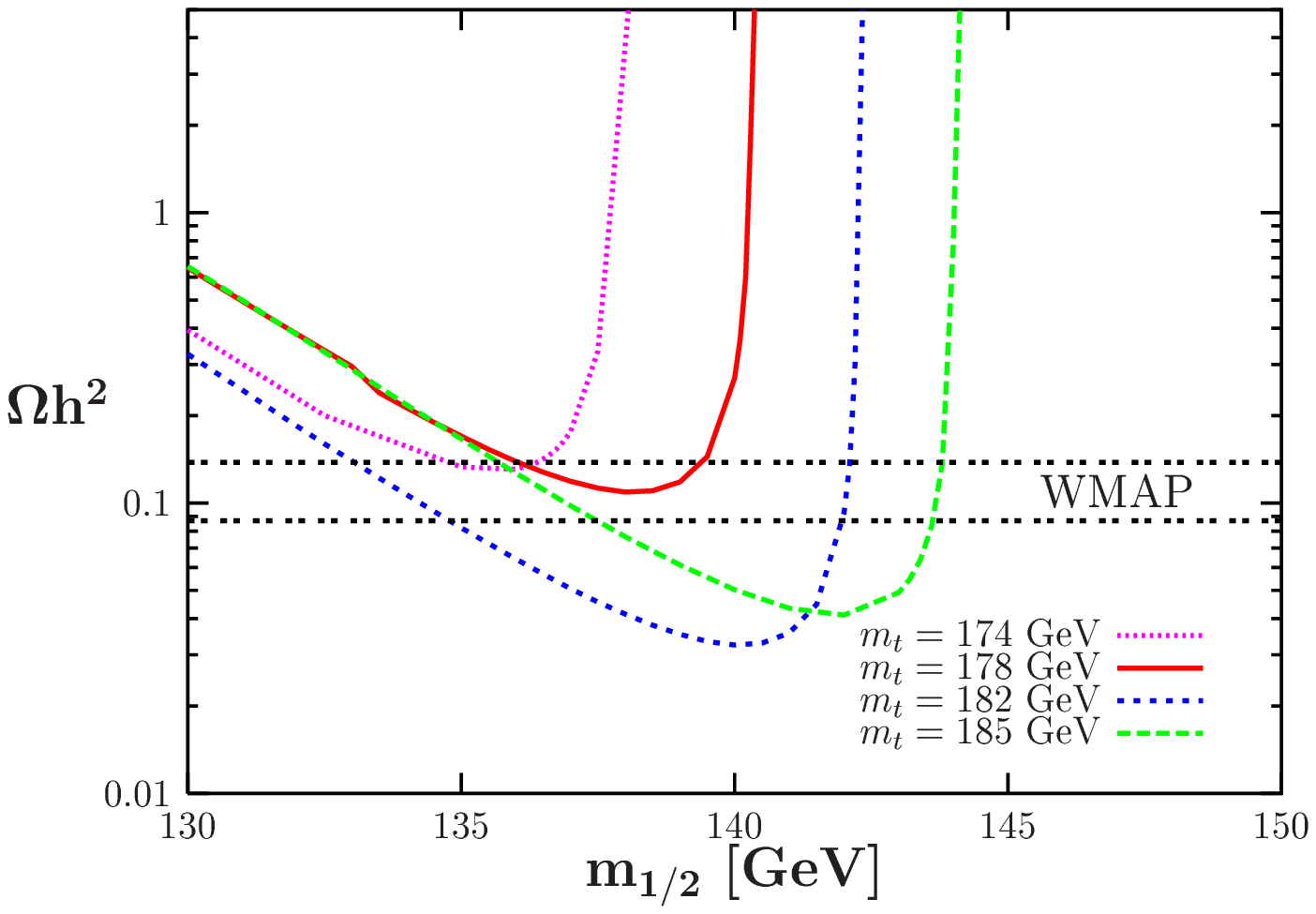,width=18cm}}
\end{center}
\vspace*{-14.8cm}
\caption[]{The lightest neutralino relic density $\Omega_{\tilde \chi_1^0}
  h^2$ as a function of $m_{1/2}$ in four mSUGRA scenarios: $i)$ $m_t=178$
  GeV, $m_0=1.5$ TeV, $A_0=-1$ TeV, $\tan\beta=30$ leading to $m_h \simeq 117$
  GeV; $ii)$ $m_t=182$ GeV, $m_0=-A_0 =1$ TeV, $\tan\beta=10$ leading to $m_h
  \simeq 115$ GeV; $iii)$ $m_t=185$ GeV, $m_0=1$ TeV, $A_0=0$, $\tan\beta=20$
  leading to $m_h \simeq 116$ GeV; $iv)$ $m_t = 174$ GeV, $m_0 = -A_0 =1.5$
  TeV, $\tan\beta = 30$ leading to $m_h \simeq 116$ GeV.  In all cases $\mu>0$
  is assumed. The area favored by WMAP, $0.087 \leq \Omega_{\tilde \chi_1^0}
  h^2 \leq 0.138$, is also indicated.}
\vspace*{-.1cm}
\end{figure}

The origin of these two branches can be understood from Fig.~2, where we plot
the scaled relic density as a function of $m_{1/2}$ for several combinations
of the remaining parameters. Recall that the relic density is basically
inversely proportional to the thermally averaged $\tilde \chi_1^0$
annihilation cross section. This cross section reaches its maximum for $\mlsp$
very close to, but just below, $m_h/2$. It remains quite large for somewhat
smaller $\mlsp$, since the finite kinetic energy of the LSPs still enables
them to annihilate resonantly; note that decoupling occurs at temperature $T
\simeq \mlsp/20$, where the kinetic energy is still significant. On the other
hand, the thermally averaged cross section drops very quickly once $\mlsp >
m_h/2$, since the (positive) kinetic energy can then only move the LSPs even
further away from the $h-$pole. This explains \cite{gs} why the curves in
Fig.~2 are not symmetric around their minimum.\footnote{We note in passing
  that the usual definition of finetuning \cite{finetune} would predict that
  no finetuning of $m_{1/2}$ is required to obtain the right relic density if
  one happens to sit at the minimum of one of the curves in Fig.~2, assuming
  the value of $\Omega_{\tilde \chi_1^0} h^2$ in the minimum falls in the
  range (\ref{omrange}), since here the derivative $d (\Omega_{\tilde
    \chi_1^0} h^2)/ d m_{1/2}$ vanishes. This is rather counter--intuitive,
  given that the $h-$pole region only extend over a narrow range in
  $m_{1/2}$.}

If this minimum corresponds to a relic density below the lower limit in
(\ref{omrange}), one has two allowed ranges of $m_{1/2}$. The asymmetry of the
thermally averaged cross section implies that the range to the right of the
minimum is very narrow; indeed, the scan used for Fig.~1 often failed to find
this region. This explains the rather ragged nature of the thin yellow strip
at $m_{1/2} \simeq 142$ GeV and $m_0 < 2$ TeV. On the other hand, if the
minimum of the relic density falls in the range (\ref{omrange}), a single
allowed range of $m_{1/2}$ results. The depth of this minimum is determined
essentially by the strength of the $h \tilde \chi_1^0 \tilde \chi_1^0$
coupling. Note that this coupling requires higgsino--gaugino mixing, which
generally is suppressed if $|\mu|^2 \gg M_Z^2$. Expanding to first order in
small quantities, one finds that the coupling of the LSP, which is Bino--like
here, to the light CP--even Higgs boson scales like
\beq \label{hcoup}
g_{h \tilde \chi_1^0 \tilde \chi_1^0} \propto \frac {M_Z ( 2 \mu \cos \beta +
  M_1) } {\mu^2 - M_1^2},
\eeq
where we have assumed $\sin\beta \simeq 1$ and $M_1 \cos \beta \ll |\mu|$,
$M_1 \simeq 0.4 m_{1/2}$ being the soft breaking Bino mass. For the examples
shown in Fig.~2, $\mu \sim 400$ to 600 GeV $\gg M_1$. Eq.(\ref{hcoup}) then
shows that the $h \tilde \chi_1^0 \tilde \chi_1^0$ coupling decreases with
increasing $\tan\beta$ and increasing $|\mu|$. In turn, $|\mu|$ increases with
increasing $m_t$, increasing $|A_0|$ (if $A_0 < 0$ or $A_0 > m_{1/2}$), and
increasing $m_0$. This latter behavior explains why the two yellow strips in
Fig.~1 merge into one if $m_0 \geq 2.1$ TeV.

The dependence on $m_0$ and $A_0$ is further explored in Figs.~3. The region
where the LSP relic density falls in the range (\ref{omrange}) is again
indicated in yellow; in between these regions $\Omega_{\tilde \chi_1^0} h^2$
is too small. We again see that the (minimum of the) relic density in the
$h-$pole region falls with increasing $m_0$ and increasing $\tan\beta$,
leading to a merging of the two yellow strips at $\tan\beta=30, \, m_0 \geq
1.8$ TeV. The slope of the right yellow strip comes about since the increase
of $m_h$ with increasing $m_0$ has to be compensated by a reduction of $A_0$,
in order to keep the difference $m_h - 2 \mlsp$ approximately constant.

\begin{figure}[!ht]
{ \hspace*{.5cm} $\mathbf{m_0}$ [GeV] 
 \hspace*{6.3cm} $\mathbf{m_0}$ [GeV]} 
\begin{center}

\mbox{\epsfig{file=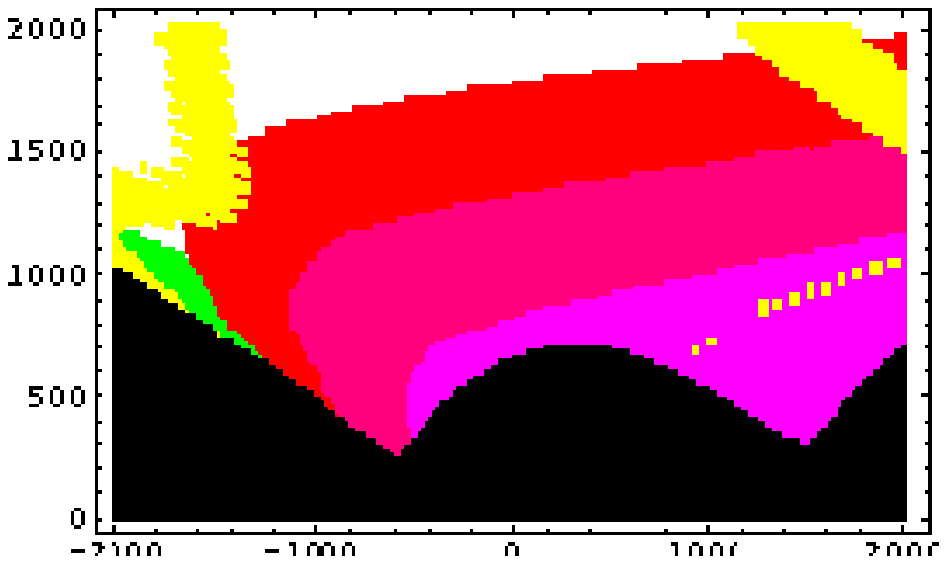,width=8cm,height=10cm} 
      \epsfig{file=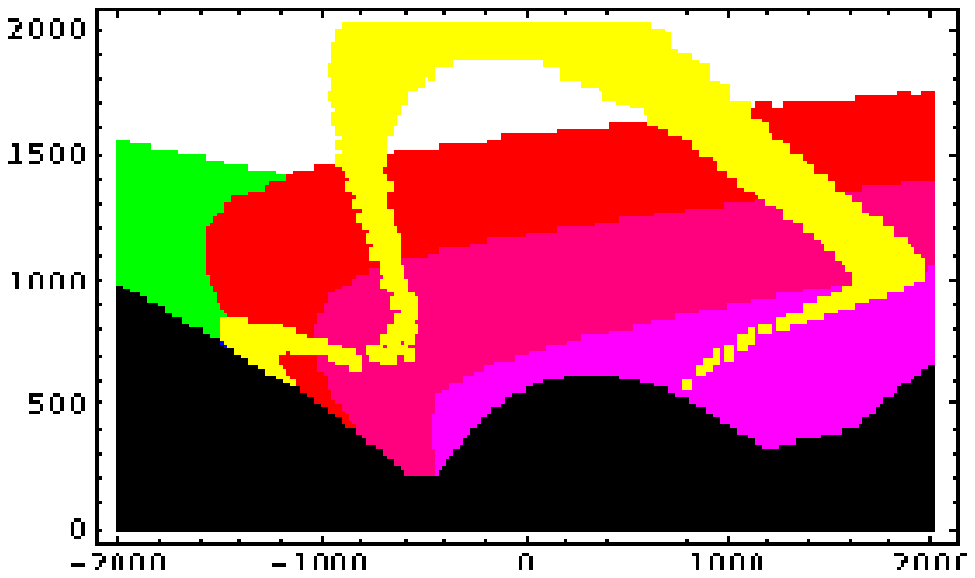,width=8cm,height=10cm} }
\end{center}
\vspace*{-.5cm}

{ \hspace*{6.cm}  $\mathbf{A_{0}}$ [GeV]
 \hspace*{5.99cm}  $\mathbf{A_{0}}$ [GeV]}
\caption[]{Constraints on the $A_0$--$m_0$ mSUGRA parameter space for 
$m_{1/2}=140$ GeV, $\mu>0$, $\tan\beta=10$ (left) and 30 (right); the top
mass is fixed to $m_t=178$ GeV. The notation is as in Fig.~1 but the $b\to s
\gamma$ constraint is partly covered by the Higgs mass constraint. Note that 
between the red and violet areas, there is an area in which 111 GeV $\lsim m_h 
\lsim 114$ GeV corresponding to the assumed 3 GeV theoretical error on the
calculation of $m_h$.} 
\vspace*{-.2cm}
\end{figure}

Note also that both $m_h$ and $\mlsp$ fall if $m_0$ is kept fixed and $A_0$ is
reduced from large negative to large positive values. Increasing $A_0$ from
large negative values means both a reduction of $L-R$ mixing in the $\tilde t$
sector and a reduction of the $\overline{\rm DR}$ top mass, both of which
reduce the corrections to $m_h$; $\tilde t$ mixing reaches a minimum at $A_0
\sim m_{1/2}$, but the corrected top mass keeps decreasing. In turn, the LSP
mass is reduced by gaugino--higgsino mixing if $\mu$ is reduced; $\mu$ also
reaches a minimum near $A_0 \sim m_{1/2}$. In addition the two--loop RGE for
$M_1$ contains a term which reduces (increases) the weak--scale Bino mass for
positive (negative) $A_0$. We see that for both $m_h$ and $\mlsp$ two effects
contribute with equal sign as $A_0$ is increased from large negative values to
$A_0 \sim m_{1/2}$, while the two effects tend to compensate each other if
$A_0$ is increased further. The overall $A_0$ dependence turns out to be
somewhat faster for the LSP mass, so that the crucial difference $m_h - 2
\mlsp$ increases from slightly negative to significantly positive values as
$A_0$ is increased. The left yellow strips in Figs.~3 therefore correspond to
the solution to the right of the minimum in Fig.~2; the extremely strong
dependence of $\Omega_{\tilde \chi_1^0}$ on $m_h - 2 \mlsp$ in this region
explains why the left strip is significantly narrower than the right one. We
note that, in addition to the two--loop terms in the RGE for the gaugino
masses, the finite (threshold) corrections to these masses \cite{bagger} are
also important here; without them, significant parts of the parameter space
shown in Figs.~3 would be excluded by the LEP chargino searches.

\section{Phenomenology of the $h-$pole region}

The results of the previous section indicate that $h-$exchange can lead to an
acceptable LSP relic density for quite wide ranges of $m_0$ and $A_0$,
whereas $m_{1/2}$ is constrained to be close to 140 GeV. In this section we
discuss the phenomenology of this region of parameter space. 

To this end we first quantify the upper and lower bounds on sparticle and
Higgs boson masses in this region. To be conservative, we again allow a 3 GeV
theoretical uncertainty in $m_h$ when interpreting the LEP Higgs search
limits. Moreover, we allow $m_t$ to lie anywhere between 171 and 185 GeV; this
corresponds to the current 90\% CL range \cite{d0top}.

In view of the theoretical uncertainty of the SM prediction for $g_\mu-2$, as
well as the strong model dependence of the prediction for $B(b \rightarrow s
\gamma)$, we performed two different scans, as shown in the second and third
column of the Table. The first scan employs our standard set of constraints,
including the requirements (\ref{amu}) and (\ref{bsg}). In contrast, the
second scan requires a positive MSSM contribution to $a_\mu$, 
\beq \label{amuee}
1.06 \cdot 10^{-9} \leq a_{\mu\,{\rm SUSY}} \leq 4.36 \cdot 10^{-9} \, ,
\eeq
corresponding to the 90\% CL allowed region when only using data from
$e^+e^-$ annihilation into hadrons for the evaluation of the SM contribution.
In the $h-$pole region scenarios with such a large $a_{\mu\,{\rm SUSY}}$ and
sufficiently heavy Higgs spectrum can be found only for $\tan\beta \gsim 15$
and not too large $m_0$. The combination of rather small squark masses and
large $\tan\beta$ leads to large $\tilde t - \tilde \chi^\pm$ loop
contributions to $B(b \rightarrow s \gamma)$. In fact, in our minimal model,
i.e. for strict squark mass universality and no flavor mixing at scale $M_X$
and a value of $V_{ts}$ essentially equal to that in the SM, the requirements
(\ref{amuee}) and (\ref{bsg}) are incompatible in the entire $h-$pole region.
In our second scan we have therefore ignored the constraint (\ref{bsg}); as
discussed in Sec.~2, this is justified if some $\tilde b - \tilde s$ mixing is
present, or if significantly different values of $|V_{ts}|$ turn out to be
allowed in the context of mSUGRA.

\begin{table}[h!]
\begin{center}
\begin{tabular}{|c||c|c|}
\hline
Quantity & Range~I & Range~II \\
\hline
 $m_{\tilde e_R} \simeq m_{\tilde \mu_R}$
 [GeV] & [708, -- ] & [299, 1300]  \\
$m_{\tilde e_L} \simeq m_{\tilde \mu_L}$ [GeV] & 
[713, -- ] & [311, 1300]  \\
$m_{\tilde \tau_1}$ [GeV] & [627, -- ]& [98.8, 934]  \\
$m_{\tilde \tau_2}$ [GeV] & [714, -- ]& [306, 1130]  \\
$m_{\tilde \nu_\tau}$ [GeV] & [708, -- ]& [281, 1130]  \\
\hline
$m_{\tilde \chi_1^\pm}$ [GeV] & [105, 122] & [105, 115]  \\
$m_{\tilde \chi_2^\pm}$ [GeV] & [295, 1820] &[297, 580]  \\
$m_{\tilde \chi_1^0}$ [GeV] & [52.9, 60.7] & [53.4, 58.4]  \\
$m_{\tilde \chi_2^0}$ [GeV] & [105, 122] & [105, 115]  \\
$m_{\tilde \chi_3^0}$ [GeV] & [280, 1820] & [280, 574]  \\
$m_{\tilde \chi_4^0}$ [GeV] & [293, 1820] & [294, 578]  \\
\hline
$m_{\tilde g}$ [GeV] & [383, 482] & [365, 433]  \\
$m_{\tilde d_R} \simeq m_{\tilde s_R}$ [GeV] & [774, -- ] & [431, 1340] \\
$m_{\tilde d_L} \simeq m_{\tilde s_L}$ [GeV] & [782, -- ] & [446, 1350] \\
$m_{\tilde b_1}$ [GeV] & [607, -- ] & [302, 920]  \\
$m_{\tilde b_2}$ [GeV] & [772, -- ] & [408, 1030]  \\
$m_{\tilde t_1}$ [GeV] & [110, -- ] & [102, 791] \\
$m_{\tilde t_2}$ [GeV] & [645, -- ] & [417, 930]  \\
\hline
$m_h$ [GeV] & [114, 122] & [114, 119] \\
$m_H$ [GeV] & [228, -- ] & [216, 825] \\
$m_{H^\pm}$ [GeV] & [246, -- ] & [234, 830]  \\
\hline
$\sigma(\tilde \chi_1^0 p \rightarrow \tilde \chi_1^0 p)$ [pb] & [$3.1 \cdot
10^{-11}, \, 1.4 \cdot 10^{-7}$] & [$6.6 \cdot 10^{-10}, \, 2.0 \cdot
10^{-7}$] \\
\hline
\end{tabular}
\vspace{.2cm}
\caption{Allowed ranges of sparticle 
and Higgs masses and of the LSP--proton scattering cross sections 
in the $h-$pole region of mSUGRA under two different sets of assumptions. 
Range~I is based on the loose $g_{\mu} - 2$ constraint (\ref{amu}) and also
uses the $b \rightarrow s \gamma$ constraint (\ref{bsg}), whereas
Range~II is based on the more aggressive $g_{\mu} - 2$ constraint (\ref{amuee})
but does not impose any constraint on $b \rightarrow  s \gamma$ decays. A dash
(--) means that values well in excess of 3 TeV are possible, which we consider
to be quite unnatural.}
\end{center}
\vspace{-.8cm}
\end{table}

We see that the first set of constraints favors a rather heavy sfermion and
Higgs spectrum (with the exception of the light CP--even scalar $h$, of
course). The reason is that large values of $m_0$ are needed to satisfy both
the LEP Higgs and $b \rightarrow s \gamma$ limits, given that $m_{1/2}$ is
quite small in the $h-$pole region. In fact, sfermion and heavy Higgs boson
masses can exceed 3 TeV, the upper end of the range we scanned; the upper
bounds on their masses are then essentially set by finetuning arguments.  The
upper bounds on the masses of the higgsino--like states $\tilde \chi_{3,4}^0$
and $\tilde \chi_2^\pm$ might also be increased if values of $m_0 > 3$ TeV are
permitted.  In contrast, the lower bound on the mass of the lighter chargino
is set by searches at LEP. Due to gaugino mass unification, this implies lower
bounds on the mass of the (Bino--like) LSP, the (Wino--like) second
neutralino, and the gluino. The upper limits on all these masses are set by
the requirement that $h-$exchange leads to an acceptable LSP relic density; we
saw in the previous Section that this is possible only if $m_{1/2}$ is around
140 GeV. This restriction, as well as the limits on $|A_0|$ that follow from
the $b \rightarrow s \gamma$ constraint, also reduce the upper
bound\footnote{The upper bound on $m_h$ is that given by SUSPECT. If we allow
  an increase by 3 GeV to reflect the theoretical uncertainty \cite{herror},
  the upper limits on $\mlsp, \, m_{\tilde \chi_1^\pm} \simeq m_{\tilde
    \chi_2^0}$ and $m_{\tilde g}$ would increase by about 1.5, 3 and 9 GeV,
  respectively.} on $m_h$ significantly, relative to general mSUGRA (without
Dark Matter constraint) \cite{ben}.

Due to the large sfermion masses, prospects of Tevatron experiments to probe
such a scenario are not very good \cite{tevsusy}. The gluino mass is above the
range that can be probed in inclusive missing $E_T$ searches at Run 2. The
cross section for $\tilde \chi_2^0 \tilde \chi_1^\pm$ production would exceed
100 fb; however, the branching ratio of $\tilde \chi_2^0 \rightarrow \tilde
\chi_1^0 \ell^+ \ell^-$ decays ($\ell = e$ or $\mu$) would be at best 6\% (for
very large $m_0$), and often smaller. Even here one would need $\sim 2$
fb$^{-1}$ of accumulated luminosity to exclude the model at 99\% CL; a
5$\sigma$ discovery would need even higher luminosity
\cite{tevsusy,baer1a,baer2}.  Similarly, with the currently foreseen
integrated luminosity, Tevatron Higgs searches might exclude some of the range
of $m_h$ shown in the Table, but a 5$\sigma$ discovery seems unlikely. The
best hope might therefore be searches for $\tilde t_1$ production, either in
pairs or from top quark decays.  However, this can only probe part of the
parameter space with $\tan\beta \lsim 5$; at larger $\tan\beta$ the lower
bound on $m_{\tilde t_1}$ lies well above 200 GeV, largely because of the $b
\rightarrow s \gamma$ constraint.

In contrast, the cross section for gluino pair production at the LHC
\cite{lhcsusy} would be guaranteed to exceed 10 pb, leading to more than
$10^5$ gluino pair events per year even at low luminosity. The cross section
for $\tilde \chi_2^0 \tilde \chi_1^\pm$ production would exceed 1 pb, so over
most of the parameter space these particles should be detectable, in
principle, both directly and in $\tilde g$ decays (a detailed analysis is,
however, required to assess to which extent this can be done).  However,
slepton searches seem hopeless even at the LHC, and the searches for heavy
Higgs bosons would be promising only for $\tan\beta \gsim 50$; note that we
didn't find any solutions with $\tan\beta > 53$, whereas values as large as 60
are allowed in other mSUGRA scenarios.  Finally, the cross section for the
production of a first generation squark together with a gluino is sizable over
much of the parameter space; however, it is not clear whether it will be
detectable on top of the large ``background'' from gluino pair production.
Moreover, detection of the heavier, higgsino--like neutralino and chargino
states would be very difficult.

In this scenario sfermions would also be too heavy to be produced at the next
linear $e^+e^-$ collider \cite{lcsusy}, now called ILC for
International Linear Collider, again with the possible exception of $\tilde
t_1$. Searches for heavy Higgs bosons would here also only be able to probe
parts of the parameter space with $\tan\beta \gsim 50$. In contrast, discovery
of $\tilde \chi_1^\pm$ pair production would be guaranteed already at
center--of--mass energy $\sqrt{s} = 300$ GeV, where associate $\tilde \chi_1^0
\tilde \chi_2^0$ production should also be detectable over much of the
parameter space. The very large cross section for $Zh$ production at this
energy might also allow to detect invisible $h \rightarrow \tilde \chi_1^0
\tilde \chi_1^0$ there; this would be a ``smoking gun'' signature for our
scenario. However, with a rough analysis using the program {\tt HDECAY}
\cite{Hdecay}, we find that in the allowed region of parameter space the 
branching ratio for this decay never exceeds 1\%. At larger $\sqrt{s}$ the 
production of the heavier chargino and neutralino states together with one 
of the light states should be feasible for a significant fraction of the 
parameter space.

We also show the range of the elastic $\tilde \chi_1^0$ proton scattering
cross section from spin--independent interactions \cite{lspp}.  This cross
section essentially determines if relic neutralinos will be detectable in
direct Dark Matter search experiments. We find cross sections well below
current sensitivity.\footnote{In the relevant region $m_0 \gg m_{1.2}$ we find
  somewhat smaller cross sections than those reported in \cite{baer3}. This is
  probably due to the somewhat larger values of $|\mu|$ predicted by SUSPECT
  relative to ISASUSY. The precise calculation of $|\mu|$ in this region of
  parameter space is notoriously difficult \cite{compare}.} The upper end of
this range can be probed in the near future; note that our LSP mass is close
to the value where current experiments have maximal sensitivity (for given
cross section).  Unfortunately such a relatively large cross section is only
possible at large $\tan\beta$; for example, in the region $\tan\beta \leq 30$,
we find $\sigma(\tilde \chi_1^0 p \rightarrow \tilde \chi_1^0 p) \leq 1.3
\cdot 10^{-9}$ pb, which is close to the limit of sensitivity even for
experiments of the next--to--next generation \cite{dmsearch}.

The third column in the Table shows that ignoring the constraint on $B(b
\rightarrow s \gamma)$ dramatically reduces the lower bounds on sfermion
masses in the $h-$pole region. The reason is that much larger values of
$|A_0|$ are now allowed, constrained mostly by the required absence of
weak--scale minima of the scalar potential where charge and/or color are
broken \cite{ccb}, leading to a much reduced lower bound on $m_0$. At the same
time, the requirement of a significant positive contribution to $a_\mu$ from
sparticle loops imposes significant {\em upper} bounds on the masses of all
superparticles and Higgs bosons.\footnote{In a general MSSM the requirement of
  a positive, nonzero $a_{\mu \ {\rm SUSY}}$ imposes upper bounds on both
  gaugino and slepton masses \cite{amubound}. In the context of mSUGRA this
  translates into upper bounds on both $m_0$ and $m_{1/2}$, and hence on the
  masses of all new (s)particles. Of course, these constraints became
  strengthened when focusing on the $h-$pole region of interest to us.} The
upper bound on $m_0$ also leads to a reduced upper bound on $m_h$. This leads
to a reduction of the upper bounds on the masses of all gaugino--like states;
the reduced loop corrections to these masses, due to reduced sfermion masses,
also play a role here.

Unfortunately the reduced slepton masses tend to {\em reduce} the leptonic
branching ratios of $\tilde \chi_2^0$, making the detection of $\tilde
\chi_1^\pm \tilde \chi_2^0$ production at hadron colliders even more
difficult. On the other hand, for some part of the parameter space slepton
pair production will now be possible at the ILC; squark production should be
detectable quite easily at the LHC; and the associate production of a light,
gaugino--like neutralino or chargino together with a heavy, higgsino--like
state should also be detectable over the entire allowed parameter space at the
second stage of the ILC operating at $\sqrt{s} \gsim 750$ GeV. Note also that
the lower bound on the LSP--proton scattering cross section has increased by
more than a factor 20, relative to the range derived from the first set of
constraints. 

\section{Summary and Conclusions}

We have discussed the $h-$pole region as cosmologically viable region of
mSUGRA parameter space, in addition to the bulk, $\tilde \tau$
co--annihilation, focus point and $A-$pole regions. Here $\tilde \chi_1^0$ is
the LSP and annihilates efficiently through the exchange of a nearly on--shell
light CP--even Higgs boson $h$. The resurrection of this region is due to a
larger central value for the top mass, as well as the calculation of
additional loop corrections to $m_h$ and to the masses of the light
gaugino--like states.

We saw that this region extends over large ranges of $m_0$ and $A_0$, but
requires $m_{1/2} < 145$ GeV. This leads to strong upper limits on the masses
of all gaugino--like sparticles, in particular $m_{\tilde \chi_1^\pm} \simeq
m_{\tilde \chi_2^0} < 125$ GeV and $m_{\tilde g} < 500$ GeV. Discovery of some
superparticles would therefore be trivial at the LHC or at the ILC; 
the best bet for the Tevatron would be searches for
light $\tilde t_1$ production, which can however only probe a small fraction
of the allowed parameter space. Moreover, sfermions, higgsino--like charginos
and neutralinos as well as the heavy Higgs bosons could all be beyond the
reach even of CLIC, if the supersymmetric contribution to the anomalous
magnetic moment of the muon is small, as indicated by SM predictions for this
quantity based on $\tau$ decay data. On the other hand, if a sizable
supersymmetric contribution is required, as indicated by SM analyses based on
$e^+e^- \rightarrow$ hadrons data only, squarks and the higgsino--like states
should also be in easy striking range of the LHC and the second stage of the
ILC, respectively.

We conclude that the $h-$pole region will soon be covered by sparticle
searches at the LHC. This distinguishes it from the $A-$pole, focus
point/hyperbolical branch and $\tilde \tau$ co--annihilation regions, which
are difficult to probe comprehensively at the LHC or ILC. If light charginos
and gluinos as well as a light $h$ boson are found at the LHC, one will have
to measure $\mlsp$ and $m_h$ very precisely, in order to check whether
$h-$exchange can indeed lead to the correct $\tilde \chi_1^0$ relic density;some information on the $h \tilde \chi_1^0 \tilde \chi_1^0$ coupling will also
be required. These measurements will probably be difficult at the LHC, but
should be straightforward at the ILC.

\subsection*{Acknowledgments}
We thank Francesca Borzumati for useful discussion on $b \rightarrow s \gamma$
decays in supersymmetric extensions of the SM. MD thanks the ENTApP network
funded by the European Union under the ILIAS initiative for partial support.

\end{document}